# THE BRAN LUMINOSITY DETECTORS FOR THE LHC

H.S. Matis, M. Placidi, A. Ratti, W. C. Turner, Lawrence Berkeley National Laboratory, Berkeley, CA 94720, United States

E. Bravin, 1211 Geneva 23, CERN, Switzerland

R. Miyamoto, European Spallation Source, ESS AB, P.O. Box 176, SE-221 00 Lund, Sweden

*Abstract*

This paper describes the several phases which led, from the conceptual design, prototyping, construction and tests with beam, to the installation and operation of the BRAN (**B**eam **RA**te of **N**eutrals) relative luminosity monitors for the LHC. The detectors have been operating since 2009 to contribute, optimize and maintain the accelerator performance in the two high luminosity interaction regions (IR), the IR1 (ATLAS) and the IR5 (CMS). The devices are gas ionization chambers installed inside a neutral particle absorber 140 m away from the Interaction Points in IR1 and IR5 and monitor the energy deposited by electromagnetic showers produced by high-energy neutral particles from the collisions. The detectors have the capability to resolve the bunch-by-bunch luminosity at the 40 MHz bunch rate, as well as to survive the extreme level of radiation during the nominal LHC operation. The devices have operated since the early commissioning phase of the accelerator over a broad range of luminosities reaching $1.4 \times 10^{34}$ cm$^{-2}$ s$^{-1}$ with a peak pileup of 45 events per bunch crossing. Even though the nominal design luminosity of the LHC has been exceeded, the BRAN is operating well.

After describing the multiple applications that the BRAN can be used to monitor the luminosity of the accelerator, we discuss the technical choices that led to its construction and the different tests performed prior to the installation in two IRs of the LHC. Performance simulations are presented together with operational results obtained during *p-p* operations, including runs at 40 MHz bunch rate, *Pb-Pb* operations and *p-Pb* operations.

## INTRODUCTION

This paper describes the BRAN (**B**eam **RA**te of **N**eutrals) segmented ionization chambers installed at CERN's Large Hadron Collider (LHC) [1]. These chambers, proposed in 1998 [2], are designed to provide a tool for the optimization of the LHC's luminosity. Earlier, a similar concept was suggested for the SSC [3]. The BRAN detectors are fast ionization chambers designed to monitor the relative luminosity at LHC at the interaction regions IR1 (ATLAS Experiment) and IR5 (CMS Experiment) by sampling the shower energy produced by forward neutrons and photons from *p-p*, *p-Pb* and *Pb-Pb* collisions [4]-[9]. This paper describes the design criteria and fabrication of the BRAN, and documents comparisons between simulated and actual BRAN performance. In addition, comparisons are made with complementary luminosity information from the ATLAS and CMS experiments.

## LUMINOSITY CONSIDERATIONS

The Luminosity at each Interaction Point can be expressed as the sum of the bunch-by-bunch luminosities $L_{sb}$ from the $b_i$ pair of colliding bunches with population $N_1$, $N_2$ in Beam1 and Beam2 over the number of pairs of bunches $k_b$ colliding at the $i^{th}$ IP [10]:

$$L(IP_i) = \sum_{1}^{k_b} L_{sb}(b_i). \qquad (1)$$

For transverse Gaussian bunch distributions, the bunch-by-bunch Luminosity at the $i^{th}$ IP from the $b^{th}$ pair (*m,n*) of bunches in Beam1 and Beam2 has the general expression:

$$L_{sb}(b_i) = f_{rev}\left(\frac{N_1 N_2}{2\pi \Sigma_x \Sigma_y}\right)_{b_i} \cdot F_1(\theta_i) \cdot F_2(z_{b_i}^{os}) \qquad (2)$$

where $\Sigma_x$, $\Sigma_y$ are the convoluted transverse bunch sizes:

$$\Sigma_z(b_i) = \left[\sqrt{(\sigma_{1z}^*)^2 + (\sigma_{2z}^*)^2}\right]_{b_i} = \left[\sqrt{\varepsilon_{1z}\beta_{1z}^* + \varepsilon_{2z}\beta_{2z}^*}\right]_{b_i} \quad (z = x, y). \qquad (3)$$

The geometric luminosity reduction factor $F_1$ quantifies the luminosity reduction in crossing angle operation [11]. Deviations from the nominal crossing angle due to closed orbit distortions driven by long-range beam-beam forces can affect in a time-dependent way the bunch-by-bunch Luminosities. The function $F_2$ represents the Luminosity reduction factor from non-zero impact parameters (PACMAN bunches, orbit errors, etc.). Assuming the same optical properties for both beams, expression (2) reads:

$$L_{sb}(b_i) = \frac{\gamma f_{rev}}{4\pi \varepsilon_n}\left(\frac{N_1 N_2}{\sqrt{\beta_x^* \beta_y^*}}\right)_{b_i} \cdot F_1(\theta_i) \cdot F_2(z_{b_i}^{os}) \qquad (4)$$

where $\varepsilon_n=\gamma\varepsilon$ is the normalized bunch emittance.

For reference, the nominal parameters at the high luminosity interaction points [12],[13] in *p-p* collision mode of operation are collected in Table 1. (Lattice parameters are evaluated at the IP).

Table 1: Nominal LHC parameters at IP1 and IP5.

| Parameter | Symbol | *p-p* Run | Units |
|---|---|---|---|
| Beam Energy | $E$ | 7.0 | TeV |
| Lorentz factor | $\gamma$ | $7.46 \times 10^3$ | |
| Rev. freq. | $f_{rev}$ | 11.2455 | kHz |
| RF freq. | $f_{RF}$ | 400.790 | MHz |
| Collision freq. | $f_x$ | 40.790 | MHz |
| Bunch spacing | $\tau_b$ | 7.485 \| 24.951 | m \| ns |
| Bunches/beam | $k_b$ | 2808 | |
| Normalized emittance | $\varepsilon_n$ | $3.75 \times 10^{-6}$ | m |
| IP $\beta$-value | $\beta^*$ | 0.55 | m |
| Bunch sizes | $\sigma^*$ | 16.7 | μm |
| Bunch length | $\sigma_s$ | 7.55 \| 0.28 | cm \| ns |
| Protons/bunch | $N_{1,2}$ | $1.15 \times 10^{11}$ | |
| IP crossing angle | $\theta^*$ | 285 | μrad |
| $F_1$ form factor | | 0.836 | |
| Luminosity/Crossing | $L_x$ | $3.2 \times 10^{26}$ | $cm^{-2}$ |
| Bunch luminosity | $L_{sb}$ | $3.6 \times 10^{30}$ | $cm^{-2}/s$ |
| Nominal luminosity | $L$ | $1.0 \times 10^{34}$ | $cm^{-2}/s$ |

## THE BRAN DETECTORS

The LHC luminosity monitors consist of gas ionization chambers that detect minimum ionization particles (MIPs) near the maximum of the shower generated by the neutral particles emerging from collisions in the IPs in the zero degree neutral particle absorbers (TAN) installed where the LHC beam transitions from one to two beam pipes, at IP1 and IP5. Figure 1 shows the position of one of the BRAN detectors in one side of the interaction region. The information on the shower energy is proportional to the neutral particles energy flux and, therefore, to the luminosity at the IP of interest. The principle lends itself to a luminosity measurement on a bunch-by-bunch basis.

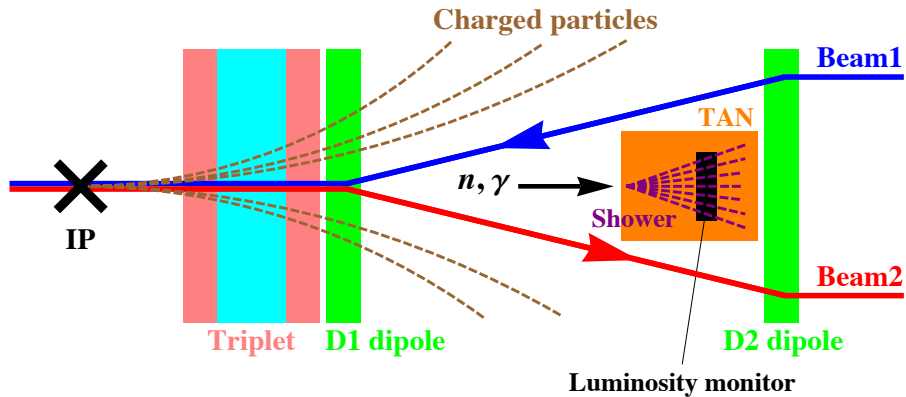

**Fig. 1:** Schematic of one of the LHC Interaction Regions.

## SYSTEM REQUIREMENTS

The system must comply with extremely stringent requirements. The speed of operation of the detector must match the 40 MHz bunch repetition rate of the LHC while standing extremely high radiation doses (up to 180 MGy/y at nominal luminosity). The requirements for the LHC luminosity monitors, specified in [14] call for a relative luminosity signal stable to 1%, a bunch-by-bunch measurement capability, crossing angle monitoring and 'reasonable integration times', where several tens of seconds is considered acceptable.

*Mechanical Design*

The specified requirements dictate the choice of a small gap, segmented gas ionization chamber built entirely out of copper, ceramic and stainless steel. The chamber has the ability to slowly flow the gas so that fresh medium is always available for the ionization process. A gas with no organic compounds was selected so that the interaction with the beam produces minimal contamination. Figure 2 shows an exploded view of the detector, which is eventually mounted into a sealed ceramic housing.

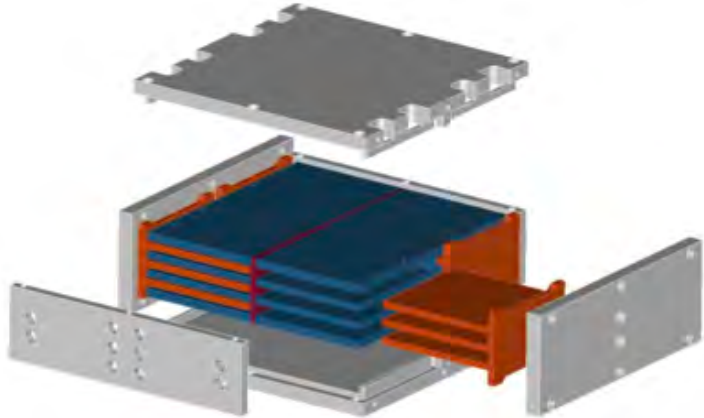

**Fig. 2**: Exploded view of the detector showing the ceramic housing and the six gaps copper structure segmented in four quadrants.

The number of gaps and their thickness have been chosen trading off between signal strength, proportional to the number of parallel gaps, and speed, which is negatively affected by the larger capacitance shown by the parallel gaps. The final compromise, supported by MAGBOLZ modeling, has been to use 6 gaps, 1 mm each, allowing for bunch-by-bunch monitoring at the nominal LHC operation. Table 2 summarizes the baseline parameters of the detectors.

**Table 2**: Ionization chamber design parameters

| Quadrant area | 1600 mm$^2$ |
|---|---|
| Gap between plates | 1.0 mm |
| Number of gaps in parallel | 6 |
| Gas composition | 94% Ar + 6% N$_2$ |
| Gas Pressure | 6 bar |
| Ionizing pairs | 58.32 mip-mm |
| E/P | 200 V/mm-bar |
| Gap voltage | 1200 V |
| Electron drift velocity | 45.0 mm/µs |
| Amplifier/Shaper gain | 0.16 µV/e$^-$ |
| rms noise | 1.24 mV |

To allow for crossing angle monitoring each chamber is segmented into four electrically isolated quadrants, so that a measurement of transverse position of the shower is possible. As the ionization chamber is designed to operate with the pressure in the range from 1 to 10 bar, a pressure vessel encloses the full detector. The ionization chamber is designed with a single ground plane electrode in a four-quadrant comb structure. Four independent signal combs interleaved with the ground plane comb provide six 1-mm gaps each. A precision-machined housing made of mica-glass composite structurally supports the ground plane comb and signal combs. Each chamber is mounted onto the end of a 605 mm long copper bar with a stainless steel end flange, and subsequently inserted into a stainless steel pressure case. The complete housing matches the instrumentation slot width in the TAN and the length of a single copper bar absorber as shown in Figure 2. The electrical and gas feedthroughs are supported at the top flange.

The BRAN ionization chambers use a 94% Ar + 6% N$_2$ gas mixture running at a nominal 6 bar pressure. This combination provides a high electron drift velocity without making use of organic molecules, helpful to avoid radiation-induced polymerization. The gas pressure controls the charge yield from the detector for a given shower intensity and the HV bias on the chamber plates is adjusted to maintain the electron drift speed near the maximum. Eqn. 5 gives the relationship between the applied electric field and gas pressure.

$$E/p_{gas} = 200\,V/(mm \cdot bar). \qquad (5)$$

At the LHC design luminosity (shown in Table 1), we simulated using the Mars Cascade code [15] that a shower population of 6,800 minimum ionizing particles (MIPs) per crossing will hit the ionization chamber. At the 6 bar nominal working pressure, the required HV bias on the chamber plates is 1.2 kV. This shower, calculated by the Mars code, produces, at the operating gas pressure, a mean ionization electron charge of 2.4 × 10$^6$ electrons. The gas mixture flows to the ionization chamber from the assembly top flange by way of a tube and holes in the ionization chamber walls. For a gas flow rate of 1 liter per hour (at 10 bar) the flow regime is laminar and the pressure drop between the inlet and outlet of the assembly is less than 15 mbar, requiring one assembly volume change per hour.

Gas is supplied to the luminosity monitor assembly from premixed gas bottles residing at the bypass tunnel area nearby the IRs. A changeover panel and a gas distribution rack provide uninterrupted gas supply and remote control of pressure and flow rate to the monitors. At the nominal flow speed of 1 l/hr, a typical sized gas bottle requires a replacement approximately once every 2 to 3 months. Although the distribution rack is about 150 meters from the monitors, the total pressure drop is less than 0.7 mbar due to the low flow rate.

Image charges of the electrons produced in the ionization process are collected and integrated by a fast front-end amplifier, followed by a shaper and a digitizing system. This signal processing chain has been designed [16] to support the nominal 40 MHz bunch rate.

*IR Installation*

The luminosity monitor assembly is mounted onto the TAN absorbers specified after energy deposition simulations [17],[18] using the same fixtures used by the absorber copper bars. Ceramic spacers isolate the monitor assembly, which is also plasma coated with ceramic. Figure 3 shows a model rendition of the detector housing and front-end electronics in the TAN. Construction layouts of the IR assembly are shown in Figure 4. The BRAN detector is positioned near the shower maximum of the hadronic electromagnetic showers.

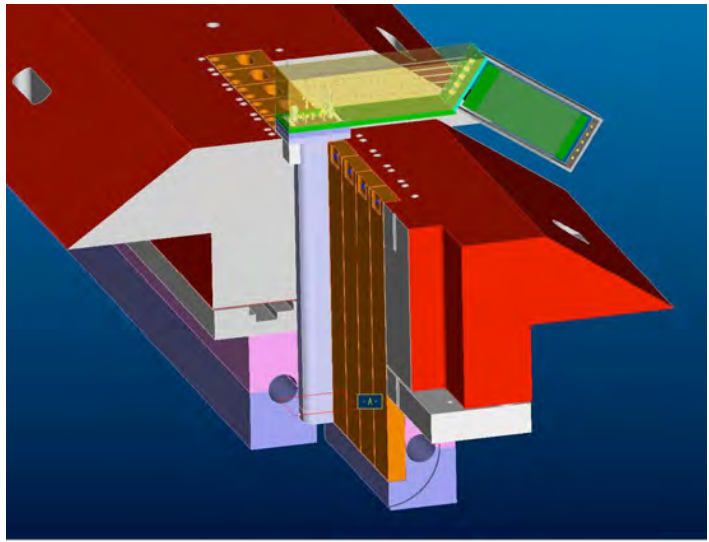

**Fig. 3:** Detail of the BRAN installation inside one of the TANs.

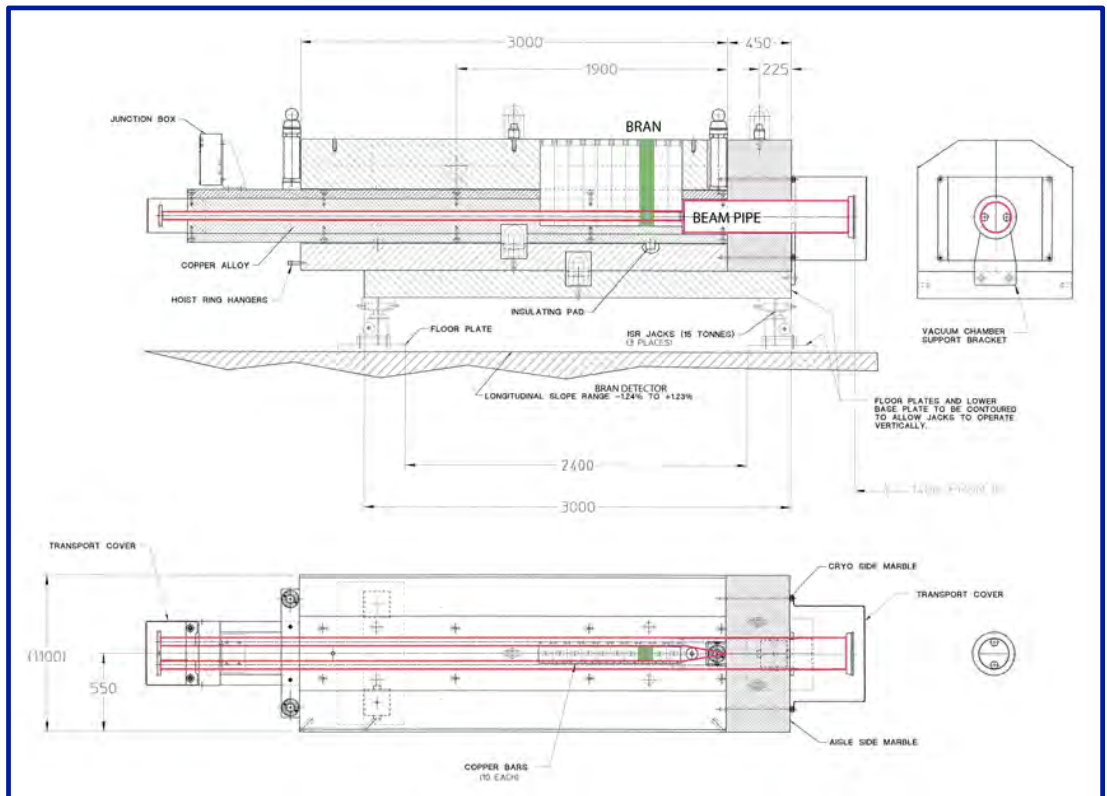

**Fig. 4:** Construction layout of the BRAN installation inside the TAN absorbers in the IR.

## ELECTRONICS DESIGN

The electronics readout comprises two separate parts, the analog processing electronics installed in the tunnel, connected to the analog shaper and the digital back-end, installed in the ATLAS and CMS service areas. The high level of radiation in the LHC tunnel requires the use of rad-hard cables between the detector quadrants and the electronics readout channels, and a careful selection of components.

### Frontend Pre-amplifiers

The main challenge of the front-end stage is to maintain a peaking time shorter than the 25 ns bunch spacing to avoid pile-up effects[19]. Since the detector signal is a current pulse, the analog front-end electronics, integrates it using a charge sensitive preamplifier with a cold-termination technique. The preamplifier is built around a classic folded cascode stage followed by a gain and buffer one to send the signal to the shaping amplifier. The time constant associated with the quadrant's capacitance and the preamplifier's input impedance is the dominant pole of the system, which is compensated in the shaper electronics.

### Shapers and Signal Distribution

The shaping amplifier's main function is to reduce the time occupancy of the preamplifier signal to less than 25 ns. The shaper's design has evolved through several iterations to a two-pole shaper, benefiting from the well-known pole-zero (PZ) cancellation scheme. The zeros cancel two poles, one associated to the detector capacitance and the preamplifier's input impedance and the other associated to the preamplifier's feedback network. The poles of the shaper implement a pseudo-Gaussian function built using a Sallen-Key low-pass filter configuration. Proper pole-zero compensation is essential in order to optimize the return of the signal to the baseline and each zero is implemented with PZ compensation capabilities. Once the signals for the individual quadrants are processed, they are buffered and distributed to the digitizers as well as to local diagnostic ports.

### Digitizers

The IBMS (Individual Bunch Measurement System) board developed by the AB/BI GROUP at CERN digitizes the analog signals. The IBMS boards have been developed as mezzanine boards for the DAB64x, a VME64x Data Acquisition Board developed by TRIUMF for LHC beam instrumentation [20]. Such a circuit board provides a very flexible platform for signal processing combining a good amount of on-board memory with the versatility and processing power of the Altera Stratix family of FPGAs. Dedicated firmware running on the DAB64x have been developed to control the instrument and to process the data. Once digitized and processed, the resulting values are shared and made available to the control room and the community through the LHC control system.

# DETECTOR APPLICATIONS

As stated in the Luminosity Specifications [12] the BRAN detectors have been conceived as relative luminosity monitors. They are used to help tune the accelerator parameters with the goal of optimizing the luminosity performance during operation without interfering with the data taking at the experiments. More specifically they are used at every store to bring beams into collision and through the beta squeeze before the start of collision production mode and the 'stable beams' used during the store.

## *LHC Operations Optimization*

Three time-dependent effects govern the time dependence of the LHC Luminosity: bunch charge decay, bunch emittance growth, and collision offsets. After the optical functions of each ring have been tuned to the design figures, substantial deviations from the design luminosities may result from charge differences in the colliding bunches at a given interaction point and from collision features (crossing angle, bunch-to-bunch overlap) specific to the orbits of bunches occupying different azimuthal positions in the rings (Pacman bunches). Information on the bunch-by-bunch luminosity contributes to bringing the beams into collision and maintaining optimum luminosity via the optimization of the bunch overlap. It also helps in identifying deviations of the optical beam parameters from the design values and therefore complements the task-oriented beam instrumentation for orbit quality control and emittance measurements.

## *Emittance Evolution*

Besides a 'direct' observation of the transverse dimensions of the beam interaction area only accessible to the experiments, a powerful tool for monitoring the time evolution of the colliding bunch combined emittance is represented by the bunch-by-bunch Specific Luminosity:

$$L_{sb}^*(b_i) \equiv \left( \frac{L_{sb}}{N_1 N_2} \right)_{b_i} = \frac{f_{rev}}{A_i^*} \cdot F_1 \cdot F_2 \ . \tag{6}$$

The interaction area at the $i^{th}$ IP reads, from (4):

$$A_i^* = 4\pi\varepsilon\sqrt{\beta_{xi}^* \beta_{yi}^*} \tag{7}$$

where $\varepsilon = \varepsilon_n/\gamma$ is the geometric emittance.

For a given crossing angle and after compensation of any unwanted collision offset (Vernier scans) the quantity (6) offers a powerful tool to record the time evolution of the interaction area and of the associated emittance growth due, for example, to Intra Beam Scattering in absence of adequate radiation damping.

The time behavior of the specific Luminosity characterizes the beam-beam physics in storage rings. It depends substantially on the nature of the colliding beams and can be quite different from the flat trend one could expect in presence of an invariable collision area. In the case of $e^+e^-$ colliders the quantity (6) exhibits a positive trend in the initial part of the operation if the beams collide at the beam-beam limit, where the bunch dimensions are larger than nominal, reaching a steady value when the bunch population decreases. In the case of *p-p* or *p-pbar* colliders the specific luminosity decreases with time providing a signature of emittance growth for one or both the colliding bunches.

The time evolution of the interaction area (7) can be monitored using the logged information on bunch charges and Luminosity. Integration times and bunch gating can be envisaged for a suitable comparison with data on the luminous regions from experiments.

An application has been implemented to display the interaction area of the proton bunches colliding in IP1 and IP5 during the physics runs. This provides a tool to help identify the contribution to the luminosity time decay originating from possible bunch emittance blow-up. Experimental results confirm the ability to characterize the emittance behavior during the store and to study possible differences among bunches in the same fill.

## *Crossing Angle Monitoring*

The information from the four quadrants of each of the BRAN Ionization Chambers can be used to evaluate and monitor the Crossing Angle at the IP1 and IP5 interaction points [21] schematically shown in Figure 5. To this purpose each of the four detectors installed at both sides of the IPs is used as a four-button position monitor to determine the transverse location of the shower via the asymmetry *(Δ/Σ)* information (up-down in IP1, right-left in IP5) from the individual quadrants. The expression

$$\theta_{x,y} \propto \frac{k_{BRAN}}{L_{sh}} (\Delta_{rl}, \Delta_{ud}) \tag{8}$$

provides the angles of the outgoing beam referred to the zero-angle reference trajectory passing through the geometrical center of the BRAN. Here $\Delta_{rl}$, $\Delta_{ud}$ are the signal amplitude differences between the right-left and up-down quadrants, $k_{BRAN}$ is a scale factor and $L_{sh}$ is the distance of the shower maximum from the IP. The total crossing angle at a given IP is the sum of the angles (8) determined from the Left and the Right BRANs. This application is intended to monitor the stability of the crossing angle rather than its actual value.

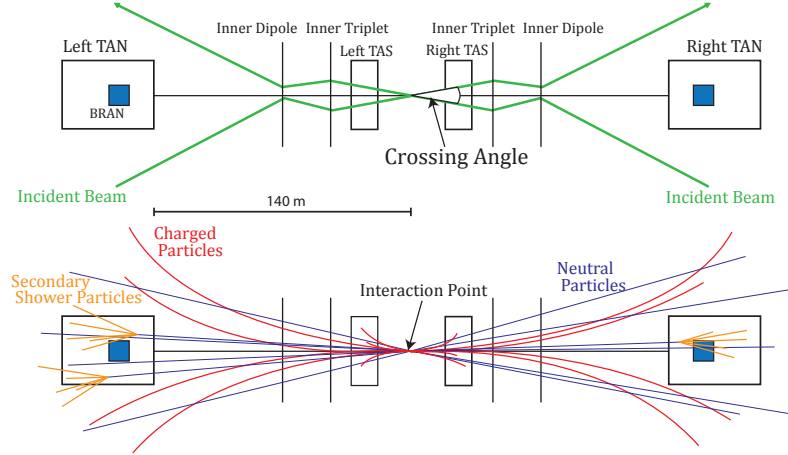

**Fig. 5**: Schematic view of one of the Interaction Regions showing the crossing angle between the colliding beams. The lower figure depicts the separation between charged and neutrals in a hypothetical particle shower resulting from an interaction.

## SYSTEM MODELING

We developed a FLUKA model [22],[23] of the BRAN detector and integrated it into the LHC's FLUKA model. The full detail of the beam line is included up to the TAN absorber, which shields the outer beam separation dipole from the forward neutral flux from the IP and houses the BRAN. Modeled TAN geometries are shown in Figures 6 and 7. Detailed FLUKA simulations have been carried on for the BRAN luminosity monitor from 3.5 to 7.0 TeV at both IR1 and IR5 for p-p collisions. They can be used to predict the behavior of the BRAN as the LHC advances to its design luminosity and energy.

Using the DPMJET [24] option of FLUKA, we have simulated the *p-p, p-Pb* and *Pb-Pb* reactions over the expected operating range of the LHC. Figure 8 shows the shower formation for *p-p* and *Pb-Pb*. The simulations [25],[26] show how the energy is deposited in the detector as a function of the beam energy and crossing angle.

Figure 8 shows the relative amount of energy in the BRAN for different combinations of colliding particles. When the LHC collides *p* on *Pb*, the signal strength from the BRAN is close to a factor of twenty higher on the side that the *Pb* beam approaches and is in good agreement [27] with the observations from the BRAN reported in the next section. Simulation results for the IP1 crossing angle (Figure 9) for three beam energies [28] confirm the anticipated increase of the shower collection efficiency at higher beam energies as the shower spread becomes more concentrated. The resulting horizontal angle is consistent with zero as expected. We performed the same simulations for IR5 and the results are identical within statistical fluctuations. A simulated comparison of the amount of energy deposited in the detector in four different modes of operation is shown in Figure 10. Here each pair of bars refers to one of the four-collider modes and shows the response at the IP1 and IP5 detectors.

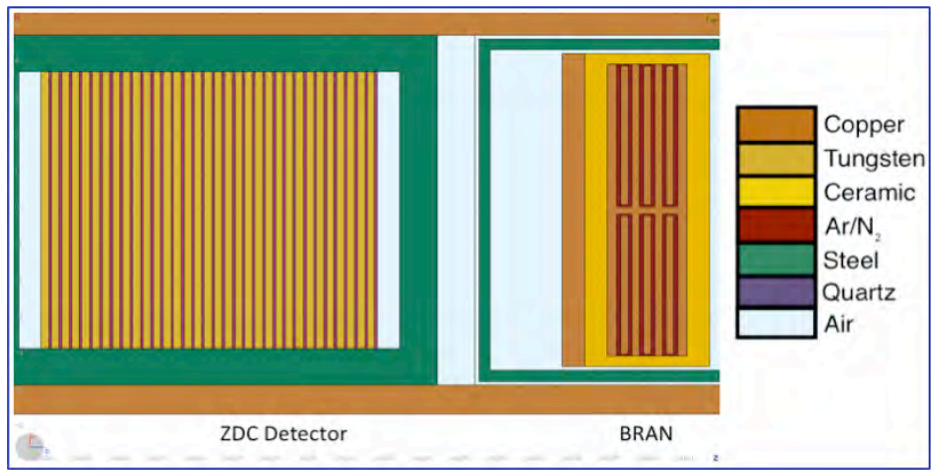

**Fig. 6:** FLUKA model of a BRAN unit installed in the TAN slot behind the ZDC detector during the 2011 LHC run.

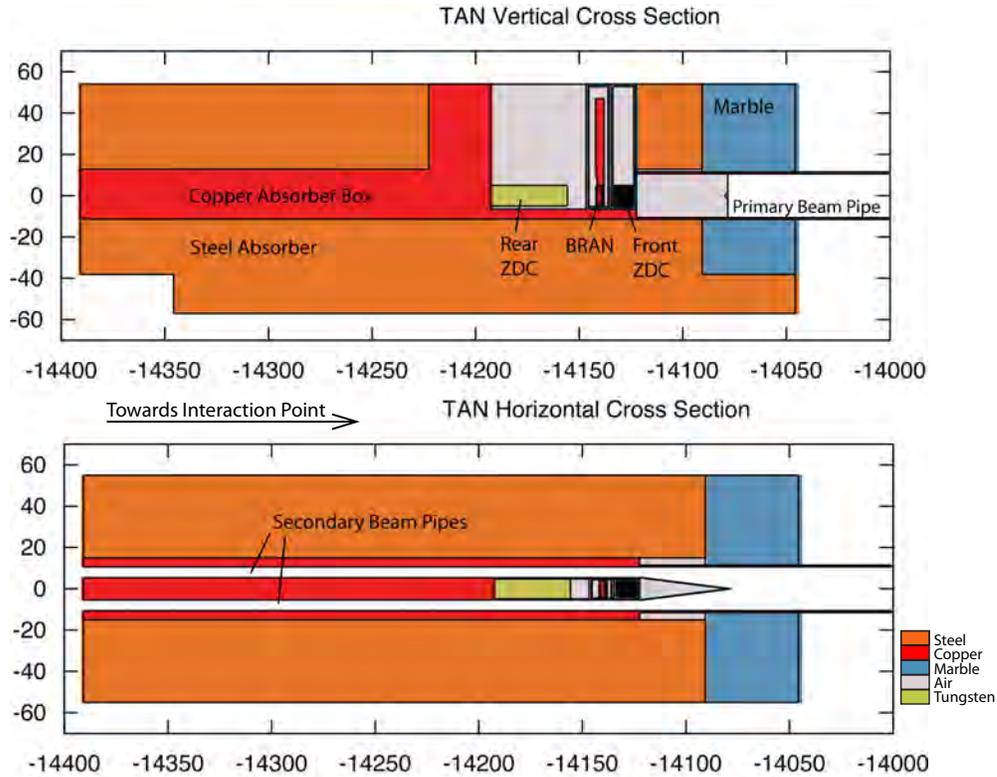

**Fig. 7:** Top and side view of the BRAN. The color white indicates the region of vacuum.

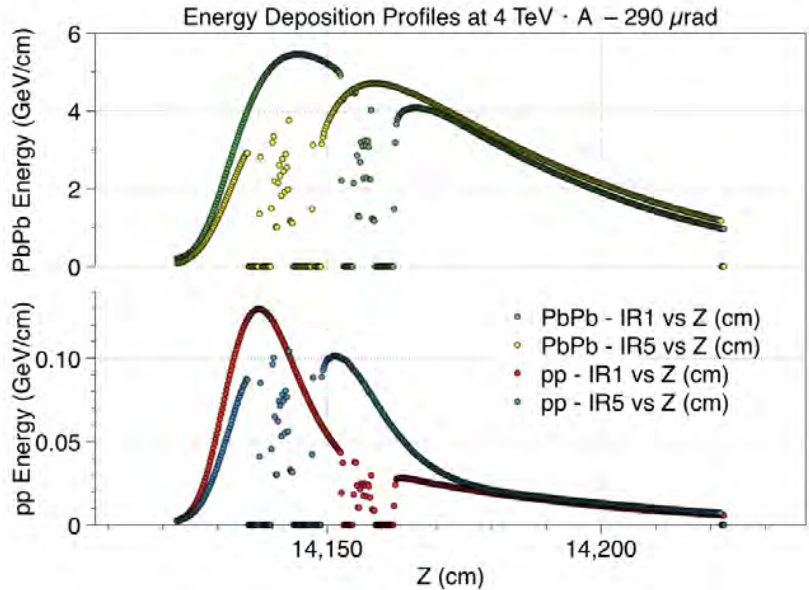

**Fig. 8**: Simulation of the TAN energy deposition for *p-p* and *Pb-Pb* collisions at 4 TeV·A. In each IR the BRAN is located in a different location, so there is an asymmetry in the shower depositions between the two IRs. The horizontal scale is the distance from the IP. The coordinate, Z, is the distance from the collision point of the IR.

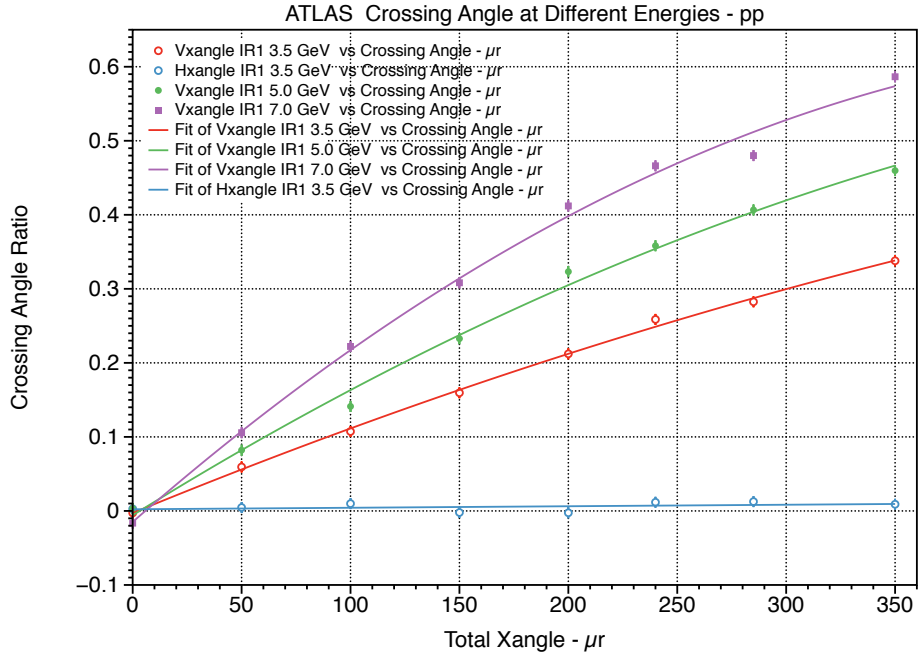

**Fig. 9**: Simulation of crossing angle ratios for the ATLAS BRAN for three LHC beam energies in *p-p* operation as a function of total crossing angle.

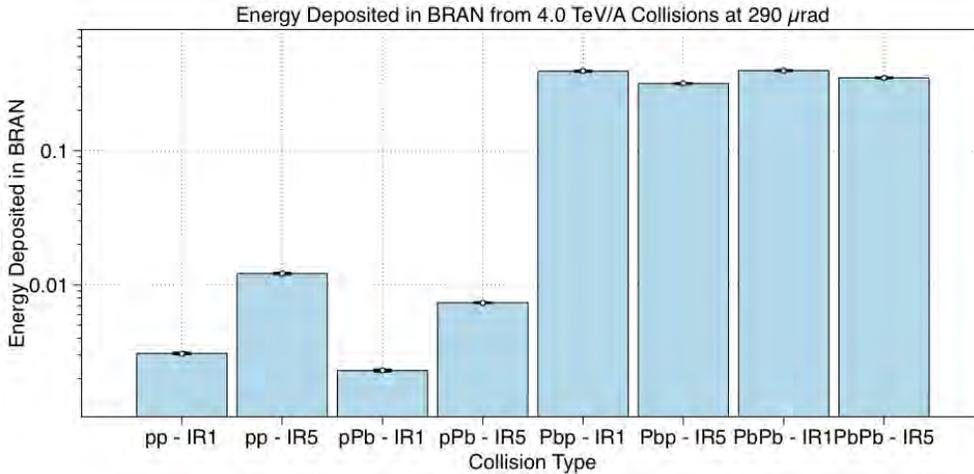

**Fig. 10:** Comparison of the amount of energy deposited in the detector with *p-p*, *p-Pb*, *Pb-p* and *Pb-Pb* collision. Note, that the y-axis is a semi-logarithmic scale.

## R&D PROGRM AND TESTS

During the experimental R&D phase of the BRAN design, a prototype of this detector has been tested extensively at the Advanced Light Source (ALS) at Lawrence Berkeley National Laboratory, at RHIC at Brookhaven National Laboratory, and at CERN's SPS. We used an electron beam at the ALS to validate the conceptual design and demonstrate the 40 MHz operation. The performance of the prototype detector was then compared to that of the existing luminosity monitors at RHIC. Finally, we validated the model with a high-energy proton beam at the SPS. The results of these experiments are presented in this section. During these studies, we carefully studied the radiation properties of the materials.

*40 MHz Resolution and Signal Speed at the ALS*

The ALS, which is a synchrotron machine, was used to test the detector speed with the help of a dedicated fill pattern and an x-ray beamline. Using a special housing with a thin wall to allow photon penetration, we irradiated the chamber and compared its signal time sequence with that from a BPM [29]. Intensity scans demonstrated good linearity with no sign of saturation, even at signal amplitudes much higher than those expected at the LHC. Signal deconvolution techniques were adopted to compensate minor residual pileup. These experiments allowed us to validate the modeling results and to demonstrate the capability of full signal processing within the desired 25 ns resolution.

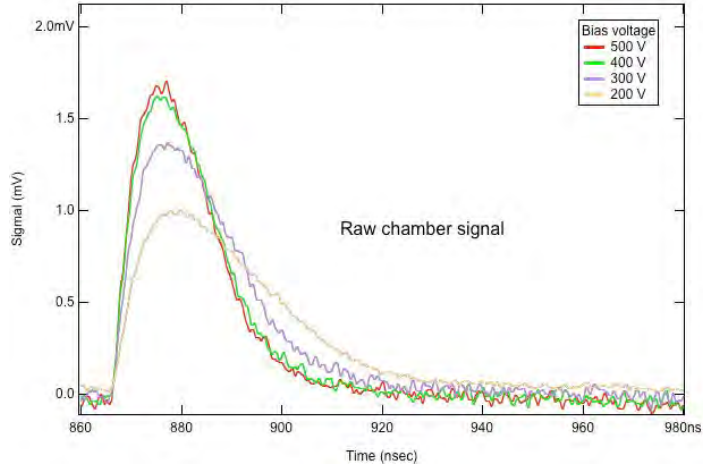

**Fig. 11**: Bias voltage scan.

Then at the ALS, we used the 1.2-1.5 GeV electron beam at the BTS (Booster-To-Storage ring) transfer line and operated it in single bunch mode at 0.5 Hz. With this setup, we could also complete a set of measurements to validate the model of the system. In particular, we found the highest detector speed by increasing the bias voltage at a fixed pressure until the drift speed reached a plateau. The time to peak height (10 ns) shown in Figure 11 demonstrates that the luminosity of adjacent bunch pairs with 25 ns separation can be resolved with E/p = 200 V/bar.

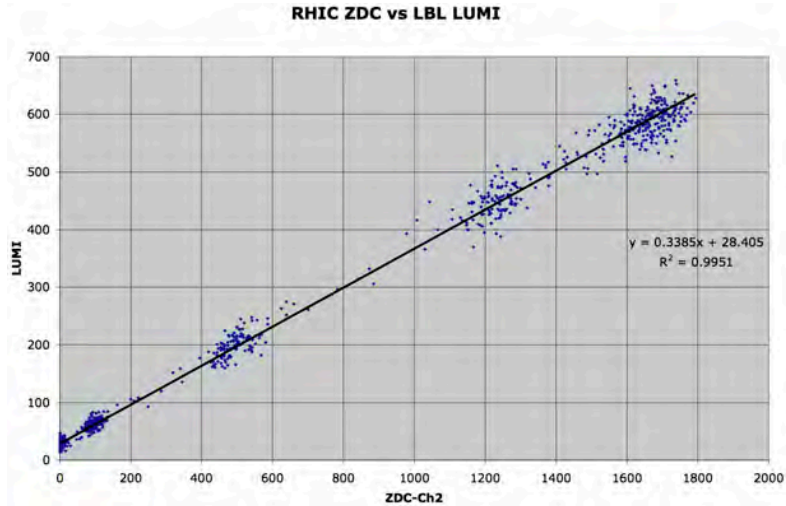

**Fig. 12**: Correlation between the counts from LUMI quadrant 2 and the RHIC ZDC.

*Rad Hardness Studies at BNL and CERN*

Due to the extreme levels of radiation, we organized two radiation damage tests, one at BNL's SPS using the RHIC Linac high intensity beam and one at the CERN ISOLDE ion source facility. The results of these tests [30] helped confirm the final choice for the materials of the detector elements.

*Tests at the SPS*

In order to validate the expected performance of the device, we had the opportunity to test the prototype BRAN detector in a collider environment and compare its performance against the ZDC detectors [31] that are used at RHIC as the primary luminosity monitors for collisions. The prototype detector was installed between two ZDC modules. The chamber was operated with a 94% Ar and 6% N gas mixture at a pressure of 8 bar. Due to the voltage rating of the prototype connectors, we operated the chamber at 400 V, which did not optimize the speed of the signal at that pressure. The discriminated analog output from all four quadrants was recorded by the RHIC control system using scalers. Figure 12 shows the correlation between one of the LUMI quadrants and the ZDC scaler during a dedicated Vernier scan. The scatter plot shows a very good linear agreement between the BRAN and the RHIC ZDC.

*Shower Studies at the SPS*

Tests at CERN's SPS North Experimental Area provided the opportunity to characterize the device and, in particular, to study the propagation of the hadronic shower in copper. This is important as the neutral products from the beam interaction at the LHC IPs produce additional showers in the TAN absorber housing the BRANs. For this measurement we used the 350 GeV extracted proton beam from the SPS against the chamber operated at a pressure of 8 bar. The system was installed on a 2-axis stage and the chamber positioned to have the beam hitting the center of a quadrant.

Digitized waveforms of each quadrant were recorded. The calculated average maxima are compared to those predicted by the Monte Carlo simulation program FLUKA in Figure 13. Each data point refers to a different copper absorber thickness. The results [32] show that the shower amplitude increases up to a 20-cm thickness of copper absorber before decreasing, in excellent agreement with the simulation.

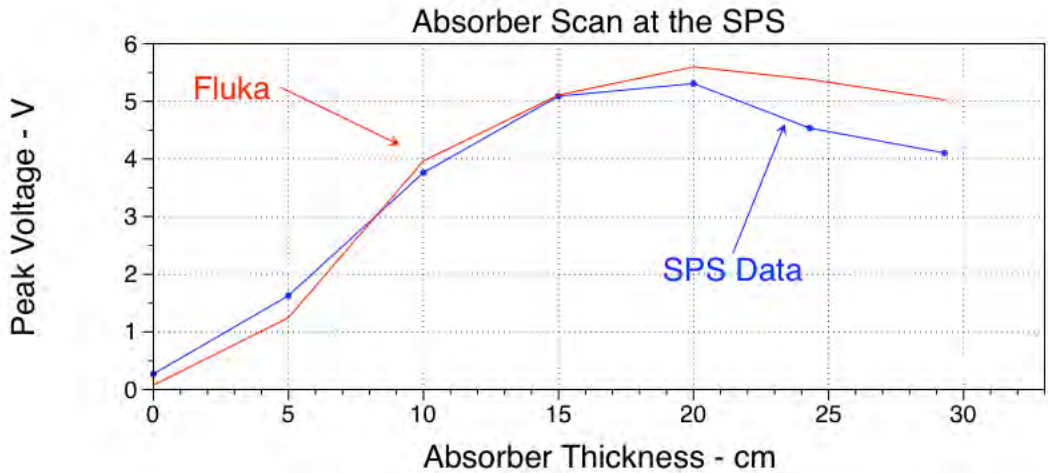

**Fig. 13**: Average energy deposited in a quadrant versus absorber thickness compared with FLUKA-predicted figures. This data was taken at CERN's SPS accelerator.

## EXPERIMENTAL RESULTS

*Modes of Operation*

The BRAN detectors are used in two different ways. In "Counting Mode" operation a hit for each bunch crossing is declared every time the voltage exceeds a threshold. This method works well at low luminosity but can saturate due to pile-up when the number of p-p interactions per bunch crossing exceeds unity. In the second method, called "Pulse Height" mode of operation, the average pulse height from each collision is recorded. The pulse height mode is linear but it tends to be dominated by system noise at low luminosity. To study these phenomena, we used the data simulated by FLUKA on an event-by-event basis. For each value of the average number of *p-p* collisions per bunch crossing, we generated the number of collisions using a Poisson distribution and summed the collected charge from one interaction. The results are plotted in Figure 14 for several thresholds. In this plot a flat curve is indicative of a linear response of the BRAN. By interpolating these results one can see the linearity for a threshold level close to about 13 mV is approximately correct in the range from 10 to 30 average collisions per bunch.

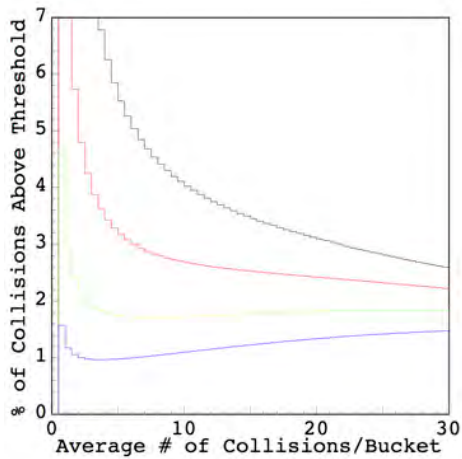

**Fig. 14**: Simulation of the fraction of collisions detected by the BRAN in counting mode as a function of the average number of collisions per bunch crossing. Thresholds are 10 mV, 15 mV, 20 mV and 25 mV from the upper curve (black) to the lower one.

*Total Luminosity in p-p Collisions*

Since the early days of the LHC commissioning, the BRAN has been a reliable tool for operating the machine and configuring the IPs for optimized collisions [33],[34]. The device has been used in every production run to bring the beams in collision and through the beta squeeze, leading to the production mode configuration of the LHC (stable beams) when the experiments publish their luminosity values. Shown in Figure 15 is a time-dependence plot of the luminosity in IP1 and IP5 recorded by the BRAN detectors and the experiments in the early phases of the LHC operation.

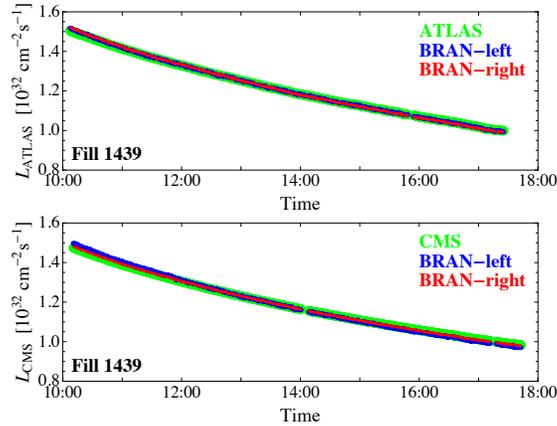

**Fig. 15**: Time dependence of the Luminosity in IP1 and IP5 recorded by the LHC experiments and the BRAN detectors.

In most of the proton runs up to LS1 [35]-[37] the LHC has operated with a 50 ns bunch spacing, an energy up to 4 TeV per beam exceeding by more than 50% the nominal bunch charge and with an emittance significantly smaller than design value, resulting in peak luminosities up to $7\times10^{33}$ cm$^{-2}$s$^{-1}$. The BRAN responded without problems to the challenge posed by the bigger amount of pileup effects resulting from the improved machine performance.

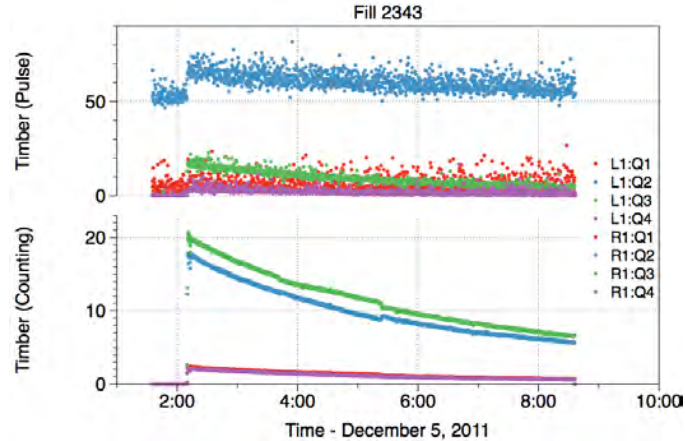

**Fig. 16:** Timber values collected for Fill 2343 for each quadrant of the IR1 BRAN detector. The top plot is for the right side in pulse height mode; the bottom plot is for the left side in counting mode.

During the 2011 run, we configured the LHC's data logging system (Timber)[*] to collect data in counting mode for the left side of an interaction region (as viewed from someone standing in the center of the LHC ring), and in pulse height mode for the right side. In addition, both modes are simultaneously read out for each BRAN detector and available for monitoring. For counting mode, the number of counts for each colliding bunch pair was accumulated for 1 s. For pulse height mode the pulse height of each colliding bunch pair was also averaged over 1 s. Figures 16 and 17 show the response from both the IR1 and IR5 BRAN detectors. The left side (operating in counting mode) has less statistical noise than the right side (which operates in pulse height mode). If the time period of the pulse mode were increased, its statistical error per measured point would decrease. This data show that most of the charge is collected in half of the ionization chamber (two quadrants) when the beams collide at a 240 µrad crossing angle. Both Figures 16 and 17 show a slight change of the energy deposition in the most populated quadrants at the time 5:20 after the beams are brought into collision. This shift in energy deposition was confirmed by data from the LHC BPM data which recorded a ~100 µm orbit shift at the same time.

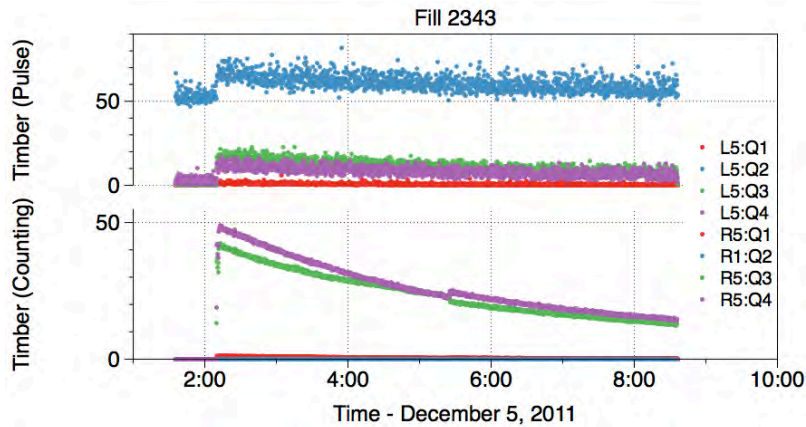

**Fig. 17**: Timber plots for the IR5 BRAN.

*Bunch-by-bunch Luminosity*

Figure 18 shows a comparison between measurements of bunch-by-bunch luminosity at IP5 recorded with the BRAN and the CMS detectors. The data was taken for one proton fill with 150 ns bunch spacing and 348 colliding bunch pairs at the beginning of the fill with a total luminosity of about $2\times10^{32}$ cm$^{-2}$s$^{-1}$. Each data point shows the average value and the standard deviation of measurements taken over 10 minutes. The agreement between the results from the two independent detectors is very good and the histogram of the relative differences for all 348-bunch pairs shown in the lower plot is consistent with a 1.0% standard deviation.

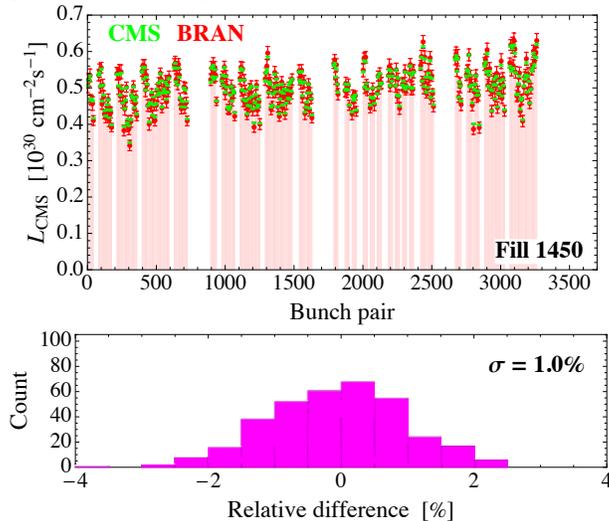

**Fig. 18**: Snapshot of bunch-by-bunch luminosity measurements at IP5 from CMS and the BRAN (upper plot), and histogram of the relative difference between the two data sets (lower plot).

The standard deviation of the relative differences remains at the same 1% level up to the end of the about 12 hours fill with a luminosity level of about $1\times10^{32}$ cm$^{-2}$s$^{-1}$. The BRANs operated in counting mode can thus provide, after proper calibration, a 1% precision level both for the total and the bunch-by-bunch luminosity.

*Monitoring 40 MHz (25 ns) Collisions*

Starting during the last weeks of the 2012 run the LHC was configured to collide trains of bunches separated by the 25 ns design time interval. This allowed both the experiments and the accelerator to test and validate their systems at the design bunch separation. Bunch-by-bunch collision data at 40 MHz rate were recorded at the BRAN detectors and is shown in Figure 19. The absence of a tail following the end of the pulse train at approximately 374 bunch pairs confirms the ability of the detector to resolve the nominal 40 MHz collision rate.

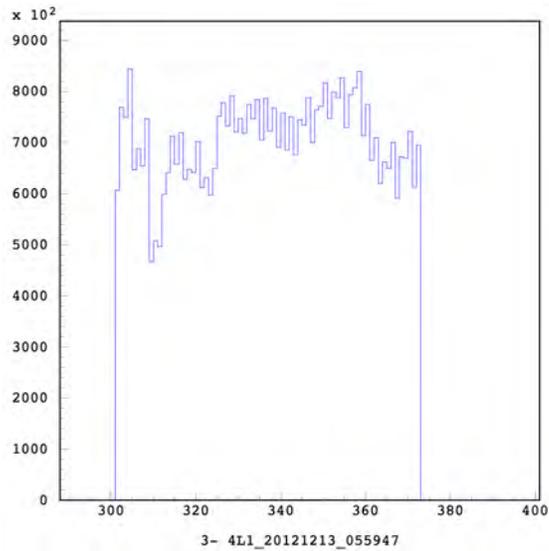

**Fig. 19**: Data collected at 40 MHz bunch collision rate. The x-axis is the collision bunch number and the y-axis shows the relative bunch intensity.

Figure 20 shows the histogram of the signal from the four quadrants for the left BRAN in IR1. Most of the signal is going into the lower two quadrants because of the vertical crossing angle in that IR. The data includes beam bunches that are empty and thus contribute to the pedestal. The pedestal is very narrow indicating that the noise is low compared to the signal. The lower curves show that the detector is not saturating and therefore can measure the full dynamic range of the shower.

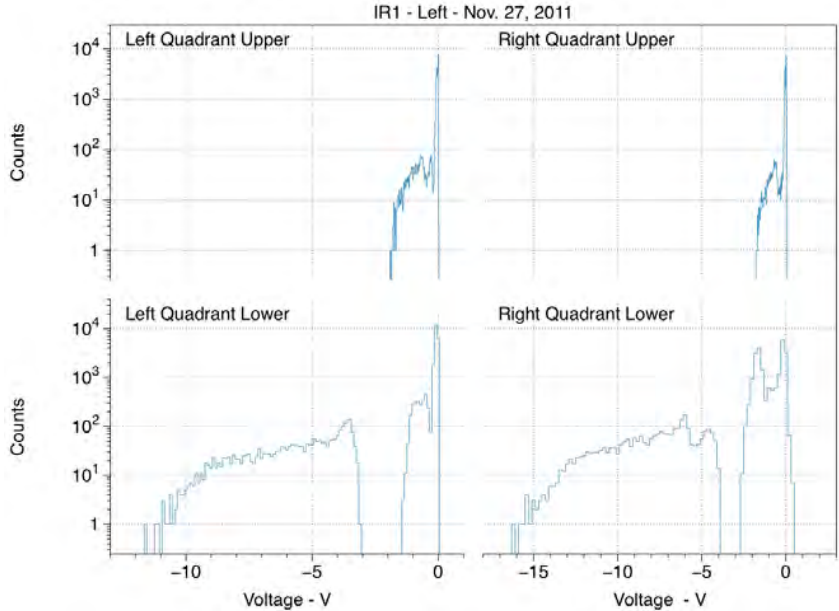

**Fig. 20**: Pulse height distribution of the analog signals from the left BRAN in IR1. The quadrants are arranged in the geometry that is viewed from the collision point in IR1.

*Comparison with the ATLAS and CMS Luminosity Detectors*

Figure 21 shows the relative difference between the luminosity data from the BRAN detectors and the detectors in IP1 and IP5 for an early *p-p* run. Besides the BRAN detector in the left side of IP5 (CMS, bottom plot), which had a noise issue, the other counters exhibit systematic errors of about ±1% in agreement with the BRAN design goals.

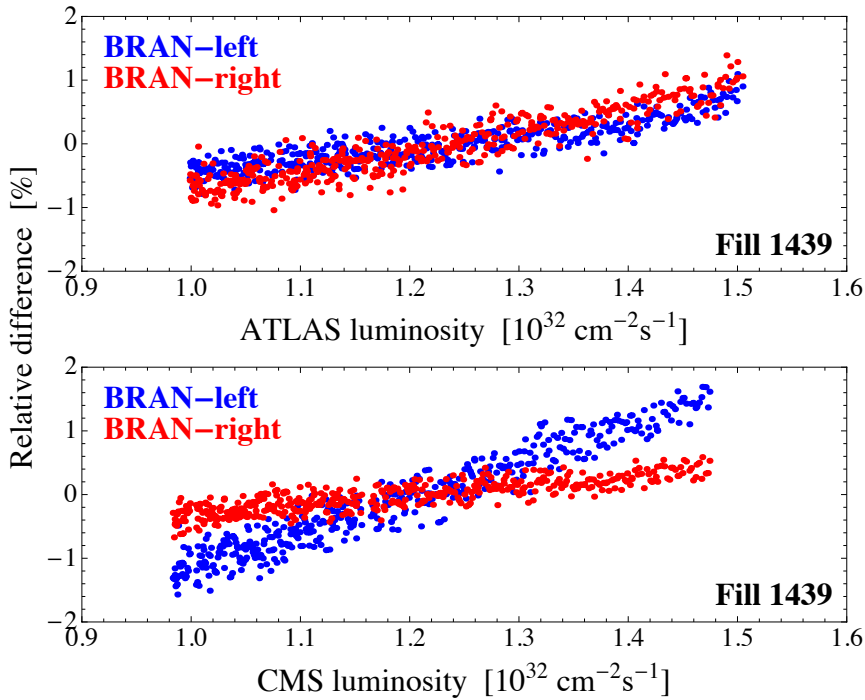

**Fig. 21**: Relative difference between the luminosity measurements from the BRAN and the ATLAS and CMS experiments in early *p-p* collision mode of operation.

A performance comparison recorded in a more recent run in 2011 is plotted in Figure 22 showing the relative difference between the rate observed at each IP1 BRAN detector versus the values provided by the ATLAS luminosity detectors. While there is a very good agreement for the left BRAN (pulse height mode, red plot) the non-zero slope from the left BRAN in counting mode is explained with saturation and electronics noise.

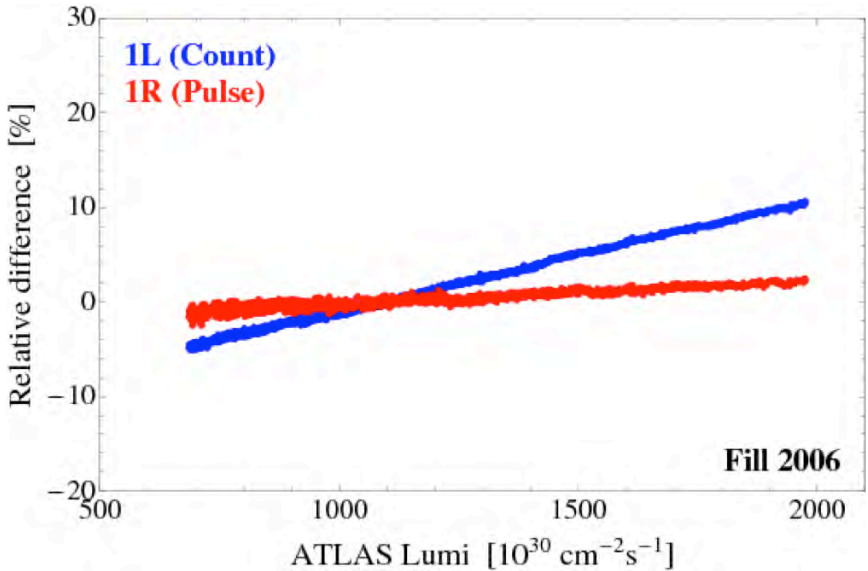

**Fig. 22**: Comparison of the relative difference between the BRAN right side (pulse height mode) and left side (counting mode) for IP1 under *p-p* operation.

Another study of the BRAN linearity is shown in Figure 23. This figure shows better linearity in pulse height mode than in counting mode as expected, with an agreement with the experiment within 1%.

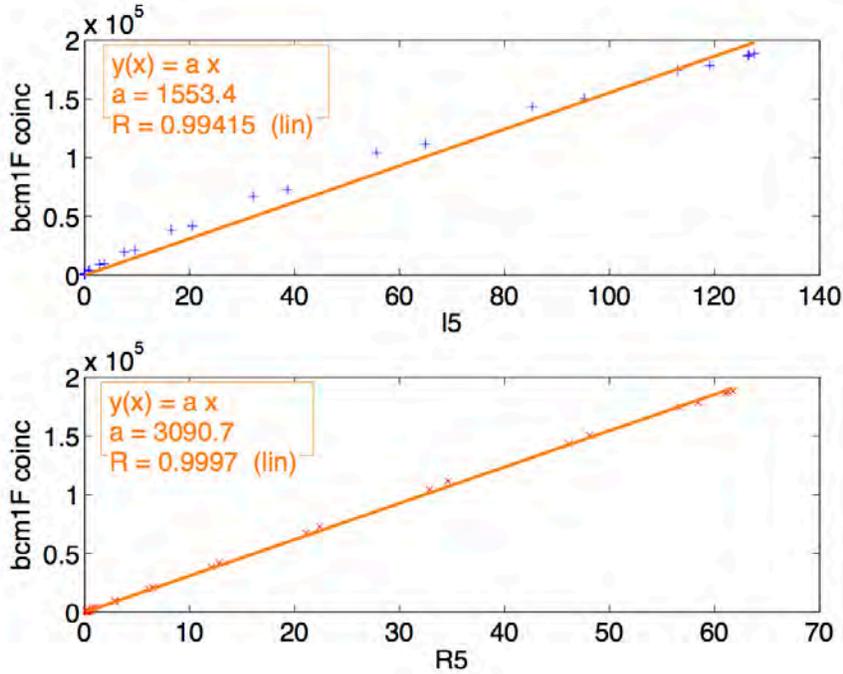

**Fig. 23**: Comparison between the performance of the Left and Right IP5 BRAN detectors (horizontal) and the CMS bcm1F coincidence detector (vertical) at different luminosities.

*Lead-lead Ion Collisions*

Although the BRANs were designed for proton operation, they are also operated routinely during lead ion collisions. The data shown here are from November 2010, when the luminosity was on the order of $10^{25}$ cm$^{-2}$ s$^{-1}$ corresponding to a collision rate of several kHz, typically two to three orders of magnitude lower than in proton operation. Figure 24 shows the relative differences between the BRANs and experiments for one lead ion fill. Compared to proton operation in Figure 21 the BRANs exhibit systematic errors about ±10 % larger.

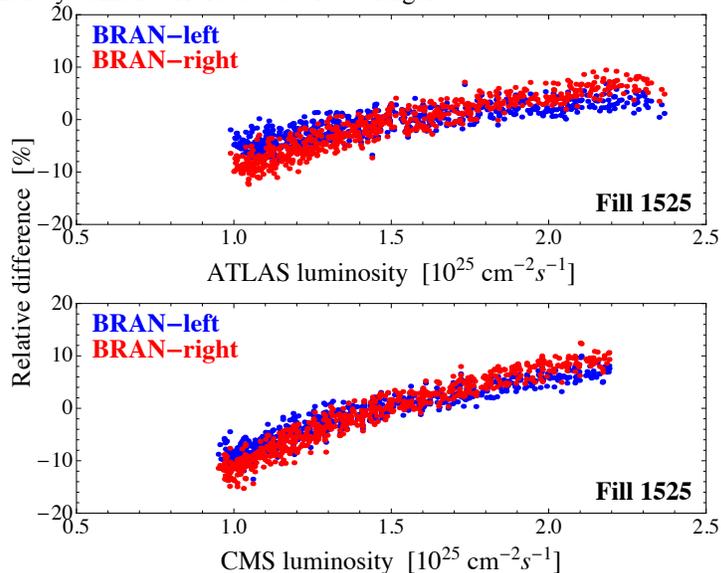

**Fig. 24**: Comparison of the relative difference between the right (pulse height mode) and the left side (counting mode) of the BRAN detector with ATLAS and CMS in *Pb-Pb* operation.

*Proton-lead Ion Collisions*

In another novel configuration, first implemented during the LHC's last run before LS1, the performance of the four detectors at both sides of the IP1 and P5 interaction regions was assessed during the accelerator operation with *p-Pb* collisions. Results collected at the even side of IP1 are shown in Figures 25 and 26 for both the pulse height and counting modes of operation. The asymmetry of the distributions generated by the vertical crossing angle is clearly visible. Similar results were collected in at IP5 where the asymmetry is in the horizontal plane.

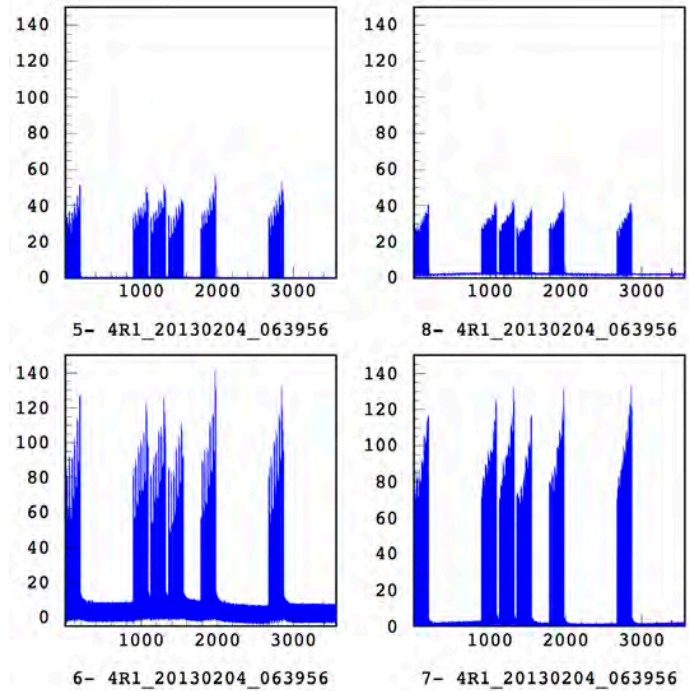

**Fig. 25**: ATLAS BRAN in pulse height mode in *p-Pb* operation. Data is from detector at right side of IP1 capturing the lead ion beam shower.

Comparisons between signals from the right and left side detectors shown in Figures 24 and 25 indicate that the shower from the *Lead* beam is approximately 18 times stronger than the corresponding shower from the *Proton* beam. Each figure shows the quadrants "as seen" from an observer situated at the beam interaction point looking toward each detector. Thus, the top two quadrants in the figures are the top quadrants of the BRAN.

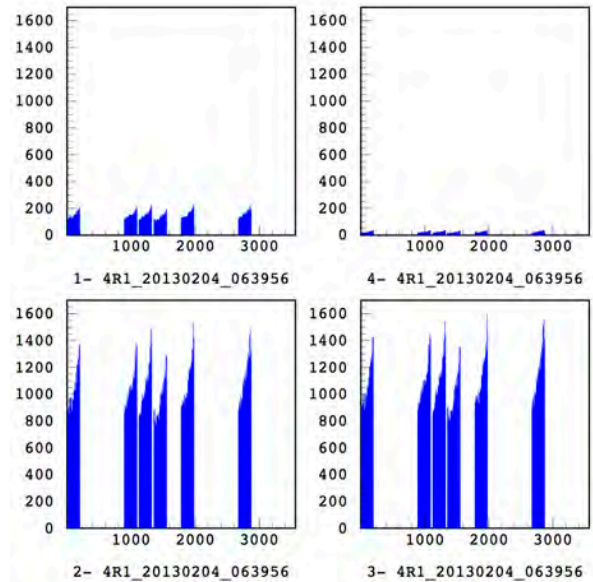

**Fig. 26**: ATLAS BRAN in counting mode, *p-Pb* operation. Data are from *left side* detector looking at the p beam. Only the bottom portion of the BRAN quadrant is significantly populated due to the vertical crossing angle.

*Van der Meer Scans*

Figure 27 shows a comparison of the performance of the BRAN detectors in IP5 with data from CMS taken during a Van der Meer Scan with the LHC operating at 4 TeV for each proton beam and a 290 µrad full crossing angle.

The response from the left detector operating in counting mode saturates as expected, while that from the right one is very linear with the CMS detector.

A Gaussian fit shows a standard deviation of (31.94 ± 0.14) µm for the right BRAN detector data as compared to a (32.06 ± 0.06) µm figure from the CMS HF detector and a (32.95 ± 0.07) µm one from the CMS BF detector. The slightly higher error in the BRAN could be improved with a longer integration time. These measurements show that the BRAN's accuracy is comparable to that of the CMS detectors.

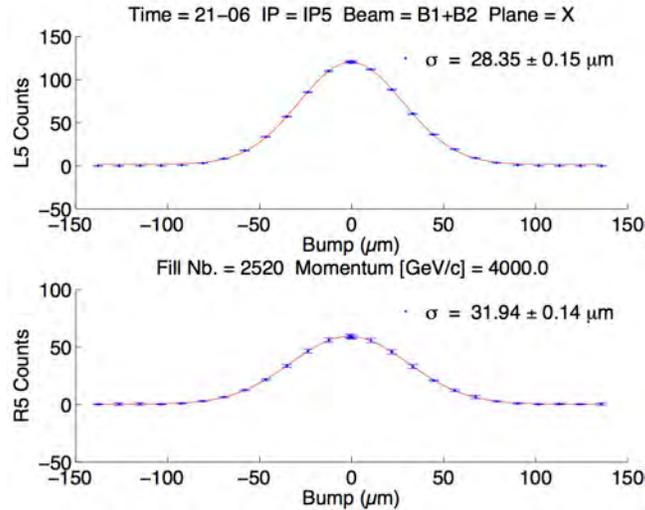

**Fig. 27**: Fit of the Van der Meer scan for the IR1 and IR5 detectors in pulse height mode. The beam energy is 4.0 TeV/beam.

*Monitoring the Crossing Angle*

Figure 28 shows a time plot of a monitoring of the crossing angle in IR1 for the 2011 LHC run with a nominal vertical crossing angle of 240 µrad. The two upper plots show the amplitude of the total shower in the four quadrants of the BRAN, proportional to the Luminosity and decreasing with time. The lower graphs show the information from the top quadrants normalized to the total shower intensity, proportional to the crossing angle. The green curve shows larger fluctuations as it comes from the R1 BRAN running in pulse height mode. The time plot also shows one instance when the ratio changes at the same time as the luminosity. Analysis of the LHC BPMs data confirmed deviations at the same.

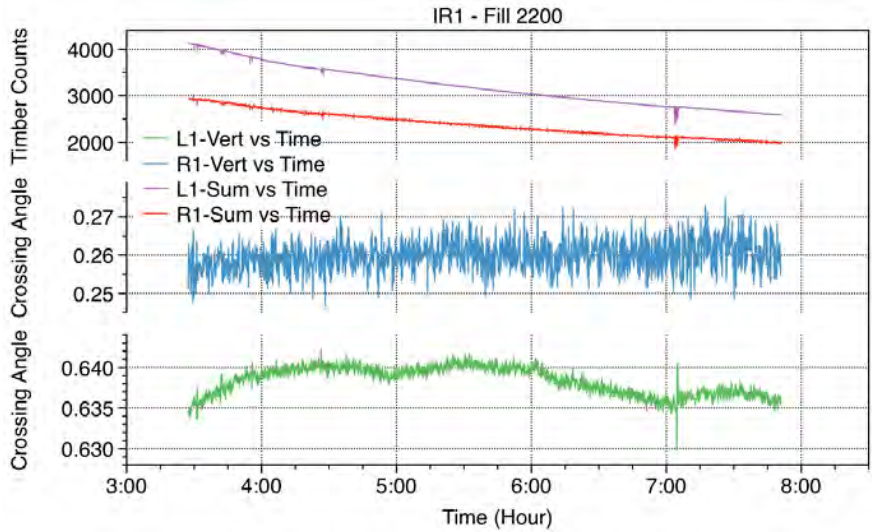

**Fig. 28**: Monitoring of the crossing angle for Fill 2200. Top curves show the time decay of the total shower intensity. The normalized bottom plots show a constant behavior of the crossing angle. Larger fluctuations from the R1 detector are due to the pulse height operation.

*Beam Property Measurements and Operator User Interface*

Figure 29 shows a 12 hours segment of Fill 1372. The BRAN data (red curve) and the Timber CMS data (green scatter plot) document the time evolution of the bunch-by-bunch luminosity and of the rms beam size of the interaction area. This was obtained by combining the BRAN luminosity information and the bunch population from the LHC beam intensity monitors via Equation (6).

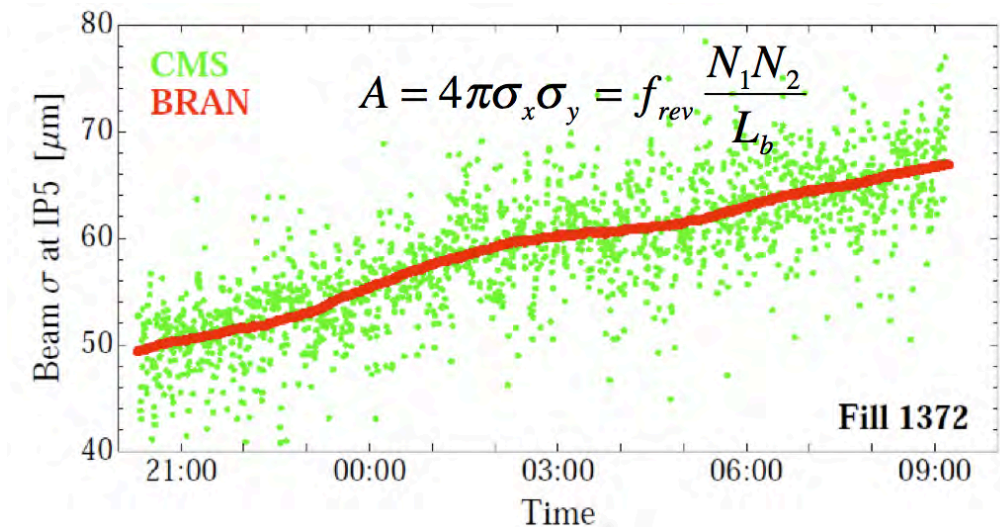

**Fig. 29**: Monitoring of the emittance growth via the time evolution of the interaction area evaluated from the specific luminosity measure at IP5 from the CMS detectors and the BRAN in pulse height mode.

The Operator Interface panel depicted in Figure 30 shows the LHC operators the information obtained by the BRAN. The two graphs on the top display the evolution of the instantaneous bunch-by-bunch luminosity in IP1 and IP5, respectively. These graphs provide information on the contribution to the luminosity time decay originating from bunch emittance blow-up. The lower four plots in Figure 30 document a batch of protons injected with larger-than-nominal emittance just after bunch 500 with the consequent loss in luminosity.

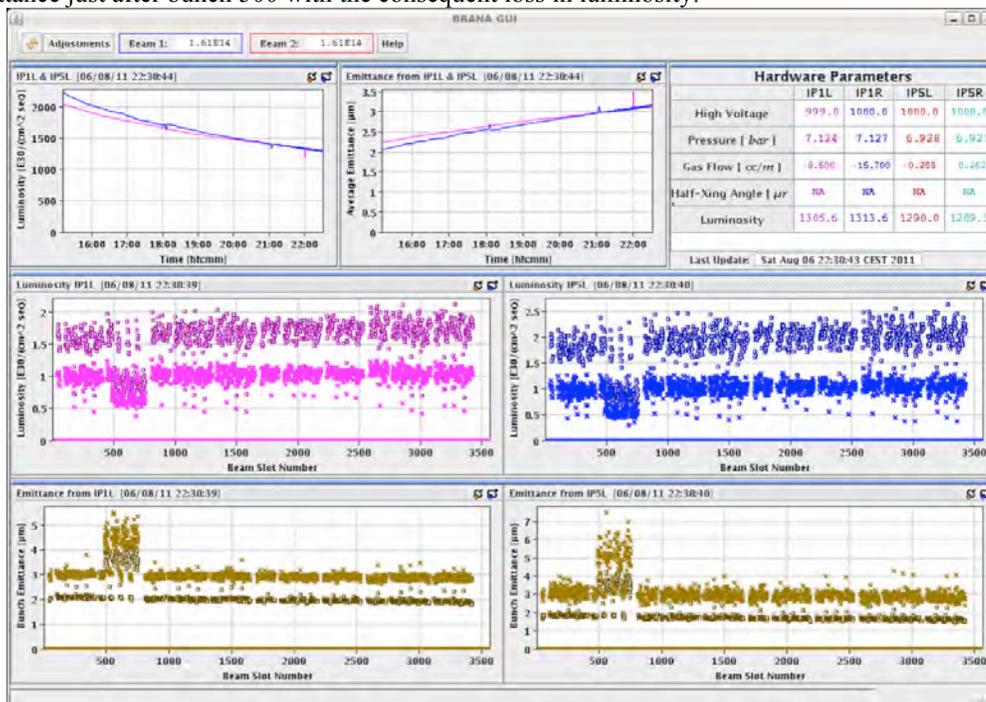

**Fig. 30:** The BRAN Operator Interface. Shown on the top left are the measurements of luminosity and the top center the measured emittance taken on August 6, 2011. The abscissa of these plots is the bunch number. They show the luminosity time decay and the emittance growth. This data in blue is from IR1, while the data in red is from IR5. The middle two plots show the luminosity in IR1 and IR5 as a function of beam slot number, while the lower two plots show the emittance as a function of beam slot number.

Among other plots captured with the BRAN, Figure 31 shows the luminosity behavior for a *Pb-Pb* run. This plot shows an excellent picture of the profile of the colliding heavy-ion beams.

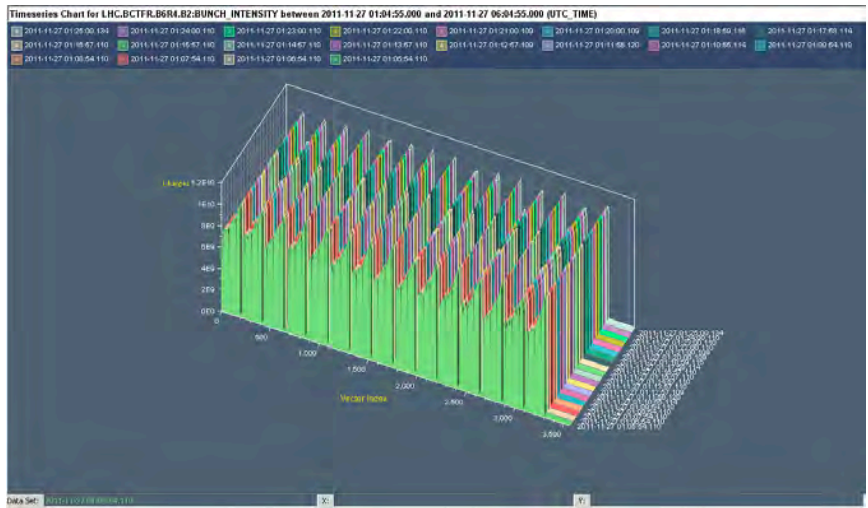

**Fig. 31**: The TIMBER plot showing the time dependence of the luminosity measured in *Pb-Pb* LHC operation for Fill 2328 on Nov. 27, 2011. On the long abscissa is the bunch number with the abort gap at the bottom right. The other axes give the time of the measurement and the bunch charge.

*The Structure of a LHC Turn*

By observing Figure 32 one can see the general structure of a turn. In the upper left graph, one sees the full turn. There are 15 bunch trains followed by the abort gap. The upper right hand graph shows two of these bunch trains followed by a third, while the lower left shows one of the bunch trains. The lower right hand graph shows that the BRAN observes counts only in one RF bucket and it does not detect collisions for the next seven buckets until protons collide at bucket nine. This information is very valuable, as it confirms the effects of intra beam scattering at injection in the SPS: when the leading bunches sit at injection while the SPS ring is getting filled, their emittance grows due to intra beam scattering. As a result, trailing bunches have higher luminosities than leading bunches.

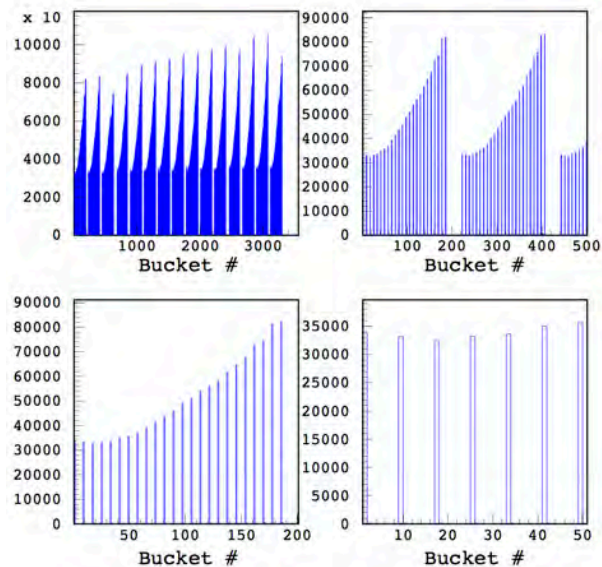

**Fig. 32**: The summed number of counts in IR5 for one turn. The difference in each panel is the bunch number. This counting mode data were taken during the PbPb 2011 run.

## CONCLUSIONS/OUTLOOK

We have described the design and the operation of a device (BRAN) that is installed in the LHC high luminosity interaction points (IR1 and IR5). This device can measure the relative bunch-by-bunch luminosity, the rms beam size and the crossing angle of the beam. The BRAN is an accelerator operations controlled device that can help operators and beam physicists diagnose the LHC and optimize its performance.

Currently the LHC's luminosity [38] reaches $1.4 \times 10^{34}$ cm$^{-2}$ s$^{-1}$ with a peak pileup of 45 events per bunch crossing. The nominal design luminosity of the LHC has been exceeded. At this intensity, the BRAN detectors continue to operate well and are routinely used by the LHC machine operators.

We have shown how by measuring the bunch-by-bunch specific luminosity it is possible to extract the corresponding values of the bunch emittances in collision, and to monitor their behavior during a beam storage cycle. We have shown that the BRAN measurements of luminosity track the luminosity measured by the ATLAS and CMS experiments. Also, the BRAN can detect subtle changes in the LHC beam.

A very important aspect of the BRAN is that it is capable of detecting fluctuations in time and on a bunch-by-bunch basis. In addition, it is the only instrument detecting beam-beam collisions during machine development studies when ATLAS and CMS are typically turned off to protect the readout systems of their electronics.

## ACKNOWLEDGEMENTS

Since the development of the BRAN luminosity detectors has taken place over almost twenty years, a large number of people have made important contributions along the way. In particular for the early stages of development we would like to thank P. Datte for his help in fabrication of early prototypes, P. Denes for finalizing the electromagnetic design and greatly simplifying the mechanical design, F. Manfredi and J.-F. Beche for their work on the low noise amplifier, D. Nygren for his help identifying Ar – N$_2$ gas mixtures as having suitable radiation hardness and electron drift speed, M. Haguenauer for help using the CERN SPS test beam facility and V. Riot for his work on the test beam data acquisition system at the SPS. A. Drees was instrumental in the successful testing of the detector in RHIC with the support of W. Fisher, and contributed the system's design. Similarly, J. Byrd contributed to the development of the system and was instrumental in testing the device at 40 MHz in the ALS at LBNL. The development at LBNL included a long list of contributors, including J. Millaud, M. Monroy, W. Ghiorso, K. Chow, H. Yaver, D. Plate, L. Doolittle, T. Stezelberger, S. Zimmermann, N. Andresen, and many others. For the later stages of development we would like to thank D. McGinnis and the LAFS group at Fermilab, in particular E. McCrory wih the help of T. Leahy (SLAC), who in collaboration with CERN's BE/CO controls group provided valuable software support. F. Cerutti of CERN provided us with a FLUKA model of the LHC Interaction Region up to the TAN and contributed many helpful discussions about the use of FLUKA. Help with FLUKA modeling came from P. Humphries, S. Hedges, and D. Nguyen. J. Stiller assisted during the simulation, construction and testing of the detector at the LHC. H. Schmickler and R. Jones supported and welcomed this effort within the BI instrument group at CERN. Lastly, we would like to acknowledge the support of the US-LHC Accelerator Project and the US-LHC Research Program, initially led by J. Strait, then S. Peggs and later E. Prebys, that together funded the design, fabrication, testing and installation of the BRAN detector.

This work was partially supported by the US Department of Energy through the US LHC Accelerator Research Program (LARP). This manuscript has been authored by an author at Lawrence Berkeley National Laboratory under Contract No. DE-AC02-05CH11231 with the U.S. Department of Energy. The U.S. Government retains, and the publisher, by accepting the article for publication, acknowledges, that the U.S. Government retains a non-exclusive, paid-up, irrevocable, world-wide license to publish or reproduce the published form of this manuscript, or allow others to do so, for U.S. Government purposes.